\documentclass{article}

\usepackage{arxiv}

\usepackage[utf8]{inputenc} 
\usepackage[T1]{fontenc}    
\usepackage{hyperref}       
\usepackage{url}            
\usepackage{booktabs}       
\usepackage{amsfonts}       
\usepackage{nicefrac}       
\usepackage{microtype}      
\usepackage{graphicx}
\usepackage{natbib}
\usepackage{doi}

\usepackage{booktabs}
\usepackage{tikz}
\usetikzlibrary{positioning,arrows.meta}

\usepackage{lineno}
\usepackage{amsmath}
\usepackage{multirow}
\usepackage{array}

\title{Task-driven sparse-view planar X-ray computed tomography for axial metrology}

\author{ \href{https://orcid.org/0009-0008-4590-8318}{\includegraphics[scale=0.06]{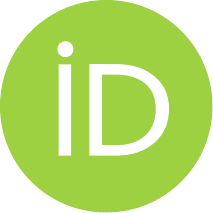}\hspace{1mm}Yibin Tian} 
    \thanks{Corresponding author.}\\
	SKL-Radio Frequency Heterogeneous Integration \\
	Shenzhen University, Shenzhen 518060, China \\
	\texttt{ybtian@szu.edu.cn} \\
	\And
	\href{https://orcid.org/0000-0000-0000-0000}{\includegraphics[scale=0.06]{orcid.pdf}\hspace{1mm}Shuming Yang} \thanks{Corresponding author.}\\
	SKL-Precision Micro-Nano Manufacturing Technology\\
	Xi'an Jiaotong University, Xi'an 710000, China \\
	\texttt{shuming.yang@mail.xjtu.edu.cn} \\
    \And
	\href{https://orcid.org/ 0000-0002-4227-8508}{\includegraphics[scale=0.06]{orcid.pdf}\hspace{1mm}Wolfgang Heidrich} \\
	Visual Computing Center\\
	King Abdullah University of Science and Technology,	Thuwal 23955-6900, Saudi Arabia \\
	\texttt{wolfgang.heidrich@kaust.edu.sa} \\
}

\renewcommand{\shorttitle}{\textit{arXiv} Task-driven XCT metrology without reconstruction}

\hypersetup{
pdftitle={A template for the arxiv style},
pdfsubject={q-bio.NC, q-bio.QM},
pdfauthor={David S.~Hippocampus, Elias D.~Striatum},
pdfkeywords={First keyword, Second keyword, More},
}

\begin{document}
\maketitle

\begin{abstract}
	Computational imaging can reduce the acquisition burden when the system is designed around task-relevant information rather than complete image recovery. We investigate this framework for sparse-view planar X-ray tomography (XCT) of layered objects, using measurement of the axial termination gap (ATG) of back-drilled printed-circuit-board (PCB) vias as a representative task. Instead of using 3D reconstruction, the proposed method estimates ATG directly from polychromatic projection data through a physics-based sensitivity model. A gated hierarchical estimator first fits ATG together with radiometric perturbations and selectively introduces global geometry and specimen-warp models only when a reduced-$\chi^2$ adequacy test and Bayesian information criterion justify the added degrees of freedom. The same low-dimensional perturbation model is incorporated into marginalized Fisher-information analyses, yielding task-specific conditional and calibration-aware Cram\'er--Rao bounds. Numerical experiments show that, under a fixed total photon budget, task-driven metrology changes little over 12–64 views, while reconstruction-driven metrology is strongly view-dependent with much larger errors. The direct estimator also follows the expected photon-limited $N_0^{-1/2}$ scaling. These results demonstrate a computational sensing strategy in which XCT acquisition, physical modeling, and inference are evaluated according to the quantitative information required by the task rather than for general-purpose volumetric image fidelity.
\end{abstract}

\keywords{Computed tomography \and 3D reconstruction \and Axial metrology \and Fisher information \and Back-drilled via}

\section{Introduction}
Computational sensing increasingly treats hardware, measurement encoding, physical forward models, and inference as components of a jointly designed system \cite{mait2018computational,sitzmann2018end,Wang2022differentiable}. This co-design perspective allows measurements to be tailored to the information required by a reconstruction or inference task, trading optical or sampling complexity for computation \cite{arguello2023deep}. Recent task-oriented approaches extend this idea further: task-driven lens design optimizes optical parameters for downstream performance \cite{yang2026task}, while latent-space imaging encodes measurements into compact task-relevant representations \cite{souza2025latent}. These studies motivate imaging systems designed for the final task rather than for faithful intermediate image formation.

Conventional X-ray computed tomography (XCT) is typically designed for volumetric reconstruction, with geometry, angular sampling, exposure, and reconstruction selected to achieve acceptable spatial resolution and artifact levels \cite{kak1988principles,feldkamp1984practical,tuy1983inversion,okawa2026snr}. Sparse sampling reduces acquisition time but makes the full-volume inverse problem increasingly ill posed. Algebraic and regularized methods, such as SART and reconstruction constrained by total variations, mitigate sparse-view artifacts \cite{andersen1984simultaneous,sidky2008image}, while learned methods seek visually faithful recovery from severe undersampling \cite{lahiri2023sparse,lin2023deep,chai2026tsv}. These approaches remain appropriate when the reconstructed image itself is the desired output, but they do not directly address how measurement number and allocation determine the uncertainty of a specified physical parameter.

This distinction is important in industrial metrology, where a 3D volume may contain millions of unknown voxels while the desired output is only one or a few quantities, such as depth, spacing, interface position, wall thickness, or defect size. Task-based imaging and statistical estimation theory indicate that the relevant quantity is the sensitivity of measurements to the parameter of interest after accounting for noise and nuisance perturbations \cite{barrett2004foundations,kay1993fundamentals}. Consequently, a measurement set may be insufficient for a stable recovery of the entire object while retaining substantial Fisher information along a particular parameter direction \cite{ly2017tutorial}. Related work in optical parameter estimation similarly emphasizes task sensitivity over global pixel fidelity \cite{barbastathis2024sensitivity}.

\begin{figure}[!h]
\centering
\includegraphics[width=0.9\linewidth]{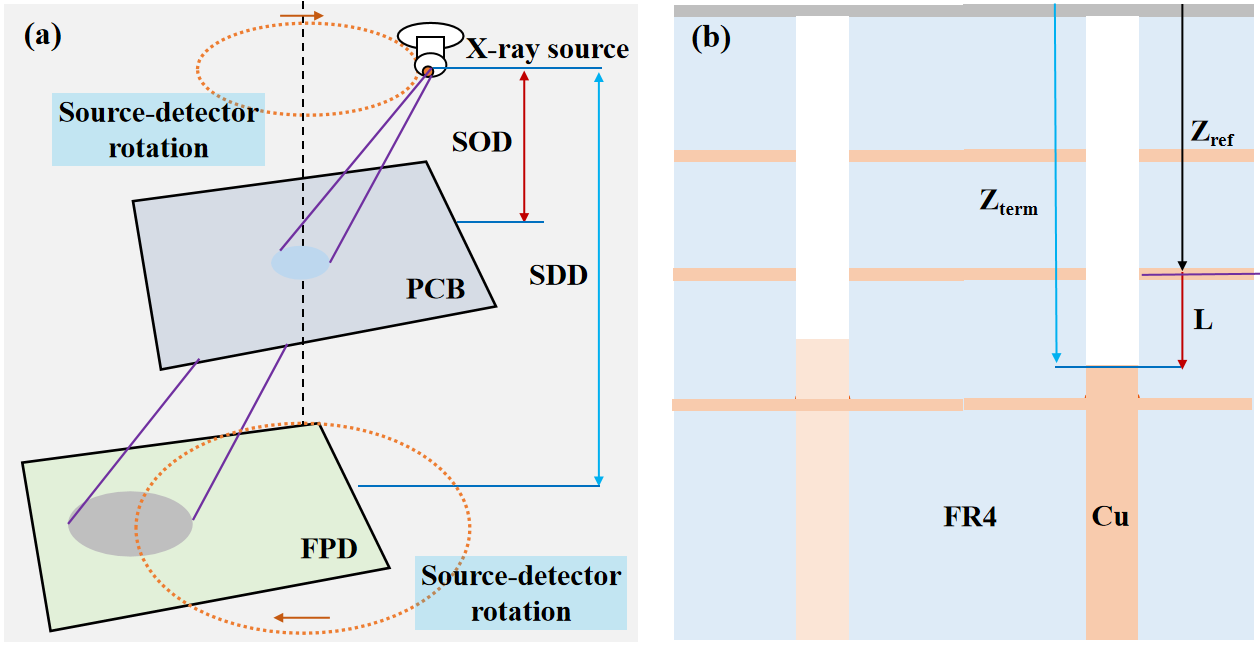}
\caption{Illustration of (a) planar XCT (SOD: source-object distance; SDD: source-detector distance; FPD: flat panel detector) and (b) a layered PCB with back-drilled vias. $Z_{\mathrm{ref}}$ and $Z_{\mathrm{term}}$ denote the axial coordinates of the reference plane and hole termination, and $L$ is the axial termination gap (ATG).}
\label{fig:planarCT}
\end{figure}

We investigate this framework using sparse-view planar XCT of layered printed circuit boards (PCBs) (Fig.~\ref{fig:planarCT}), with the axial termination gap (ATG) of a back-drilled via as representative task. The Back-drilling removes the unused conductive portion of a via, which makes the remaining axial separation important for high-speed interconnect performance. The plate-like PCB geometry naturally motivates an oblique planar XCT trajectory similar to that of computed laminography \cite{helfen2005high,marcus2022information}. This application provides a controlled test bed in which the desired output is low-dimensional, while calibration errors and board deformation can still bias axial metrology.

The proposed framework combines sparse oblique XCT acquisition, a differentiable physical projection model for direct axial metrology, and a gated estimator that expands the perturbation model only when supported by the data. ATG is estimated directly from projection measurements rather than through an intermediate 3D reconstruction, while perturbation-marginalized Fisher analyses quantify the task information preserved by a given view and photon allocation. The main contributions of this study are as follows. 

(1) a task-driven formulation of sparse-view planar XCT with projection-domain sensitivity and Fisher information for quantitative axial metrology; \\
(2) a low-dimensional hierarchical perturbation model in which radiometric effect is fitted first, followed only when necessary by global geometry and specimen warp, with Bayesian information criterion used to limit unnecessary model expansion \cite{schwarz1978estimating}; \\
(3) perturbation-marginalized Cram\'er--Rao bounds that incorporate both task sensitivity and calibration uncertainty; \\
(4) numerical comparisons of task-driven and reconstruction-driven metrology under fixed-photon-budget view sweeps, photon-budget variation, and controlled model perturbations. The central result is that task-driven and reconstruction-driven metrology can have substantially different angular-sampling requirements.

\section{Computational XCT model}

\subsection{Oblique planar XCT geometry}

We consider a stationary layered PCB centered near the origin and a cone-beam source/detector pair moving on an oblique circular trajectory (Fig.~\ref{fig:planarCT}(a)). For view $i$, the source direction is
\begin{equation}
{n}_i =
\begin{bmatrix}
\sin\alpha_i \cos\phi_i, &\sin\alpha_i \sin\phi_i, &\cos\alpha_i
\end{bmatrix}^{\mathsf T},
\end{equation}
where $\phi_i$ and $\alpha_i$ are the azimuth and obliquity relative to PCB normal. With source--object distance (SOD) and source--detector distance (SDD), the 3D position vectors of the X-ray source and the detector center are
\begin{equation}
 s_i=\mathrm{SOD}\, n_i,\quad
 d_i=-(\mathrm{SDD}-\mathrm{SOD}) n_i .
\end{equation}
The detector coordinate system comprises two orthogonal in-plane axes, $u$ and $v$, defined locally for each projection view. The detector-$v$ direction is aligned
with the projection of the PCB axial direction onto the detector plane, while the detector-$u$ direction is perpendicular to it. Thus, axial displacement produces its principal. The detector basis is chosen for all views so that the detector-$v$ direction contains the principal sensitivity to axial displacement. The system is implemented with the ASTRA Toolbox \cite{vanAarle2015astra,vanAarle2016fast}.

\subsection{Polychromatic Beer–Lambert forward model}

The expected primary photon count at the detector pixel $j$ is modeled with a discretized polychromatic Beer--Lambert law \cite{alvarez1976energy,yang2020x}
\begin{equation}
\lambda^{\mathrm{pri}}_j(\theta)
=
N_0
\sum_{e=1}^{N_E}
w_e
\exp\left[
-\sum_m
\mu_m(E_e)
\ell_{jm}(\theta)
\right],
\label{eq:poly}
\end{equation}
where $N_0$ is the incident photon budget per pixel, $w_e$ the detector-weighted spectral fraction at energy $E_e$, $\mu_m(E_e)$ the linear attenuation coefficient of the material $m$, and $\ell_{jm}$ its ray path length. The parameter vector $\theta$ contains ATG and, when activated, radiometric, geometric, and specimen-warp parameters. Cu and FR4 are modeled as two effective material classes \cite{islam2023bending}. Detector blur and focal-spot unsharpness are included following established flat-panel XCT models \cite{tilley2016model}. Scatter mismatch is represented by an additive slowly varying component that approximates the predominantly low-frequency scatter background \cite{gao2010scatter,mccann2021local}.

Raw detector measurements are modeled as Poisson-dominated counts with additive electronic read noise \cite{xu2009electronic,ding2018statistical}. A flat-field reference $F_j$ is estimated from multiple frames, followed by conventional dark- and flat-field normalization \cite{van2015dynamic},
\begin{equation}
\widehat T_j = \frac{Y_j-d}{F_j-d},
\end{equation}
where $d$ is the dark-field level. Normalized transmission is converted to a line-integral estimate using \cite{fu2016comparison}
\begin{equation}
p_j=-\log\left[\max(\widehat T_j,\epsilon)\right],
\label{eq:log}
\end{equation}
with $\epsilon=10^{-6}$. No upper clipping is applied to $\widehat T_j$; retaining samples with $\widehat T_j>1$ avoids a one-sided bias in weakly attenuated regions.

Using first-order variance propagation through the normalization and logarithmic transform, and assuming independent object and flat-field noise conditioned on the detector gain map, the post-log inverse variance is approximated by \cite{xu2009electronic}
\begin{equation}
W_j \simeq
\left[
\frac{C_j+\sigma_r^2}{(C_j-d)^2}
+
\frac{N_0+d+\sigma_r^2}{N_F N_0^2}
\right]^{-1},
\label{eq:weights}
\end{equation}
where $C_j$ is the expected raw object count, $\sigma_r$ the read-noise SD, and $N_F$ the number of flat-field frames averaged. The two terms account for object-measurement noise and finite flat-field precision, respectively; dark-field uncertainty is neglected.

\section{Direct measurement inference from sparse projections}

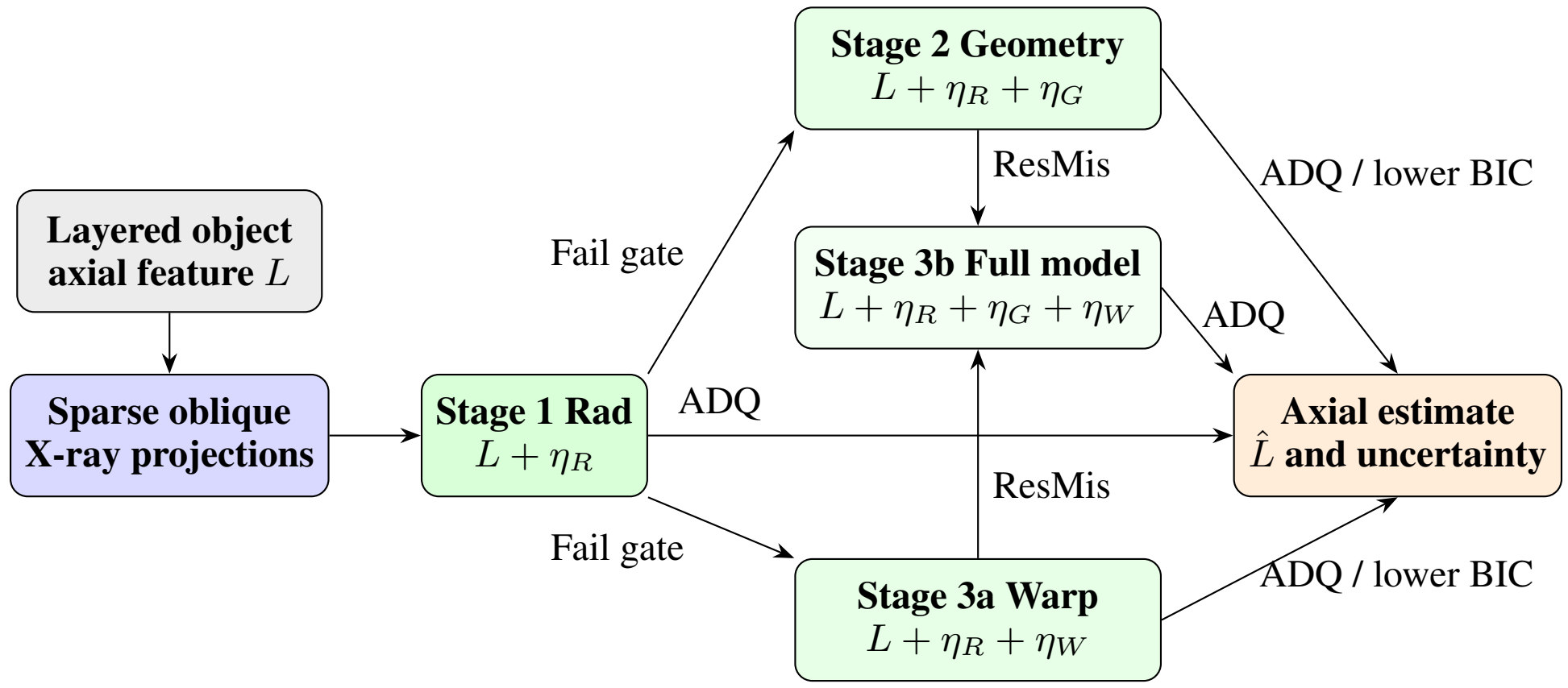
\begin{figure}[!h]
\centering
\begin{tikzpicture}[
    >=Stealth,
    node distance=0.5cm,
    every node/.style={font=\small},
    box/.style={
        draw,
        rounded corners,
        align=center,
        minimum height=1.0cm
    }
]

\node[box, fill=gray!15, minimum width=2.5cm] (obj)
{\textbf{Layered object}\\\textbf{axial feature $L$}};

\node[box, fill=blue!15, minimum width=2.5cm, below=0.5cm of obj] (proj)
{\textbf{Sparse oblique}\\\textbf{X-ray projections}};

\node[box, fill=green!15, minimum width=1.5cm, right=0.75cm of proj] (s1)
{\textbf{Stage 1 Rad}\\\textbf{$L+\eta_R$}};

\node[box, fill=green!10, minimum width=3.0cm, above right=2.0cm and 1.2cm of s1] (sg)
{\textbf{Stage 2 Geometry}\\\textbf{$L+\eta_R+\eta_G$}};

\node[box, fill=green!10, minimum width=3.0cm, below right=0.5cm and 1.2cm of s1] (sw)
{\textbf{Stage 3a Warp}\\\textbf{$L+\eta_R+\eta_W$}};

\node[box, fill=green!5, minimum width=3.0cm, above right=0.2cm and 1.2cm of s1] (sgw)
{\textbf{Stage 3b Full model} \\\textbf{$L+\eta_R+\eta_G+\eta_W$}};

\node[box, fill=orange!15, minimum width=2.5cm, right= 4.8cm of s1] (out)
{\textbf{Axial estimate}\\\textbf{$\hat{L}$ and uncertainty}};

\draw[->] (obj) -- (proj);
\draw[->] (proj) -- (s1);

\draw[->] (s1) -- node[above,pos=0.55]{ADQ} ++(2.0,0) |- (out.west);

\draw[->] (s1.north east) -- node[above left,pos=0.35]{Fail gate} (sg.south west);
\draw[->] (s1.south east) -- node[below left,pos=0.35]{Fail gate} (sw.north west);

\draw[->] (sg.east) -- node[right,pos=0.35]{ADQ / lower BIC}  (out.north);
\draw[->] (sw.east) -- node[right,pos=0.35]{ADQ / lower BIC}  (out.south);

\draw[->] (sg.south) -- node[right,pos=0.35]{ResMis} (sgw.north);
\draw[->] (sw.north) -- node[right,pos=0.35]{ResMis} (sgw.south);

\draw[->] (sgw.east) -- node[right,pos=0.35]{ADQ} (out.north west);

\end{tikzpicture}

\caption{The framework of task-driven sparse-view planar XCT axial metrology. The estimator begins with the parameter $L$ and radiometric (Rad) perturbations $\eta_R$. If this model is adequate (ADQ), the corresponding estimate is accepted. Otherwise (Fail gate or Residual mismatch (ResMis)), the hierarchy evaluates higher-complexity candidates, including geometry ($\eta_G$) and warp ($\eta_W$) candidates. If needed, a full model is considered. Among statistically adequate candidates, the lower-Bayesian Information Criterion (BIC) model is retained.}
\label{fig:concept}
\end{figure}

Fig.~\ref{fig:concept} summarizes the workflow of direct measurement of ATG in back-drilled PCB using sparse projections from planar XCT, which is detailed in this section.

\subsection{Axial gap measurement}

The quantity of interest is ATG, which is the axial separation between a known reference plane and the termination of the target via 
\begin{equation}
L = Z_{\mathrm{ref}} - Z_{\mathrm{term}},
\label{eq:axial_gap}
\end{equation}
where $Z_{\mathrm{ref}}$ and $Z_{\mathrm{term}}$ denote the axial coordinates of the reference plane and the via termination, respectively (Fig.~\ref{fig:planarCT}(b)). A conventional reconstruction-driven workflow first estimates the 3D attenuation distribution from the measured log projections
\begin{equation}
\widehat{X} = \mathcal{R}(P_{\mathrm{obj}}),
\label{eq:reconstruction}
\end{equation}
where $\mathcal{R}$ denotes the reconstruction operator, and $P_{\mathrm{obj}}$ is the combined log projections of the target object from different views. $L$ is then obtained by applying a metrology operator $M(\cdot)$ to the reconstructed volume
\begin{equation}
\widehat{L}_{\mathrm{rec}} = M(\widehat{X}).
\label{eq:reconstruction_metrology}
\end{equation}
This reconstruction-first formulation implicitly requires the intermediate volume to preserve the spatial features needed for both the localization of the reference plane and the detection of the via termination. However, in sparse-view planar XCT, these structures can have markedly different visibility and sensitivity to angular undersampling. Approximately horizontal and wide reference layers and a localized via termination therefore need not be reconstructed with comparable fidelity. As a result, global image quality measures are not necessarily reliable surrogates for axial metrology performance: a 3D reconstruction with relatively small overall image error can still exhibit a significant bias in $\widehat{L}_{\mathrm{rec}}$, whereas a volume containing conspicuous sparse-view artifacts may retain sufficient task-specific information for an accurate estimation of $L$.

Instead, We estimate $L$ directly from the projection data, without requiring an intermediate 3D reconstruction. Let $Q(L,{\eta})$ denote the predicted log-projection vector for $L$ and the perturbation parameter vector ${\eta}$. In the current iteration $(L,{\eta})$, the residual is 
\begin{equation}
 R
=
 P_{\mathrm{obj}} - Q(L,{\eta}).
\label{eq:residual}
\end{equation}
The sensitivity of the predicted measurements to $L$ is represented by the Jacobian as
\begin{equation}
 J_L
=
\frac{\partial Q}{\partial L},
\end{equation}
which is evaluated numerically using a centered finite difference
\begin{equation}
 J_L
\approx
\frac{
 Q(L+\Delta L/2,{\eta})
-
 Q(L-\Delta L/2,{\eta})
}{
\Delta L
}.
\label{eq:JL}
\end{equation}
For fixed perturbation parameters, a weighted Gauss--Newton update of $L$ is then given by
\begin{equation}
\Delta L_{\mathrm{GN}}
=
\frac{
J_L^{\mathsf T}
 W
 R
}{
 J_L^{\mathsf T}
 W
 J_L
},
\label{eq:GN1}
\end{equation}
where $W$ is the diagonal statistical weighting matrix defined from the approximate inverse variance of the log-projection measurements. The denominator is the local Fisher information-like curvature associated with $L$ when the perturbation parameters are kept fixed.

To improve computational efficiency and suppress weakly informative measurements, only detector samples with high sensitivity to $L$ are retained in each projection view. In addition, each projection view may exhibit a small constant offset in the log domain, for example, due to imperfect flat-field correction or incident-flux normalization. This offset is removed analytically by projecting the residual onto the subspace orthogonal to the constant vector under the weighting matrix $W$. Consequently, the estimator is robust to view-dependent intensity shifts without introducing an additional perturbation parameter for every projection view.

\subsection{Joint task--perturbation update}

When a perturbation model is active, the unknown parameter vector is
\begin{equation}
\boldsymbol{\theta}
=
\begin{bmatrix}
L\\
\boldsymbol{\eta}
\end{bmatrix},
\label{eq:para}
\end{equation}
where $\boldsymbol{\eta} = [\eta_1,\ldots,\eta_K]^{\mathsf T}$
contains the $K$ active perturbation parameters.. The corresponding measurement Jacobian is
\begin{equation}
J
=
\frac{\partial Q}{\partial \theta}
=
\begin{bmatrix}
J_L &
J_{\eta_1} &
\cdots &
J_{\eta_K}
\end{bmatrix},
\label{eq:joint_jacobian}
\end{equation}
with each column describing the sensitivity of the predicted log projections to one parameter.

To account for differences in units and characteristic scales, optimization is performed in scaled coordinates by normalizing each parameter with its finite-difference step. For a fixed active parameter set, weak Gaussian calibration priors are applied only to the perturbation parameters, while the task parameter $L$ remains unregularized. The Gauss--Newton update is
\begin{equation}
\left(
H_{\mathrm{data}}
+
H_{\mathrm{prior}}
\right)
\Delta{\theta}
=
G_{\mathrm{data}}
+
G_{\mathrm{prior}},
\label{eq:jointGN}
\end{equation}
with
\begin{equation}
H_{\mathrm{data}}
=
J^{\mathsf T} W J,
\qquad
G_{\mathrm{data}}
=
J^{\mathsf T} W R,
\label{eq:data_normal}
\end{equation}
where $H_{\mathrm{prior}}$ and $G_{\mathrm{prior}}$ represent the calibration-prior contributions. Their entries associated with $L$ are zero, so that no prior is imposed on the measureand. After each iteration, physical bounds and component-wise trust regions are applied to avoid nonphysical solutions and limit compensating updates among strongly coupled perturbation parameters.

\section{Gated physical perturbation model}

The nominal projection model cannot capture all the acquisition effects in a real physical system. We therefore represent departures from the calibrated system with a \emph{low-dimensional physical perturbation model}, whose parameters describe interpretable changes in source spectrum, detector response, system geometry, and specimen warp. The goal is to model dominant calibration errors that can bias ATG estimation without using unnecessary DoF that can reduce task identifiability. Because ATG, board warp, and geometric offsets can exhibit strong sensitivity coupling, perturbations are introduced hierarchically rather than fitted simultaneously. The estimator expands the active perturbation family only when supported by the data. The perturbation parameters are summarized in the Appendix (Table~\ref{tab:perturbation}).

\subsection{Stage 1: radiometric perturbation}

In the first stage, the parameter vector is
\begin{equation}
\theta_1=
[L,\eta_R]^{\mathsf T}, \quad \eta_R=
[s,\sigma_b,V_{\mathrm{kVp}},t_{\mathrm{Cu}}]^{\mathsf T}
\end{equation}
where $s$ is the scatter fraction, $\sigma_b$ is the effective blur, $V_{\mathrm{kVp}}$ controls the source-spectrum endpoint and $t_{\mathrm{Cu}}$ is Cu-equivalent filtration. These parameters account for the dominant global radiometric perturbation observed in industrial XCT systems.

\subsection{Stage 2: global geometry perturbation}

The geometry model uses
\begin{equation}
\eta_G=
[\delta_M,\Delta\alpha,\Delta v]^{\mathsf T},
\end{equation}
where $\delta_M$ is the relative magnification error,
\begin{equation}
M=M_0(1+\delta_M),
\end{equation}
$\Delta\alpha$ is a common obliquity error and $\Delta v$ is the detector translation along the local detector-$v$ axis. Independent SOD and SDD offsets are not used in the final estimator because their effects are strongly correlated. Magnification directly captures their most identifiable first-order effect.

\subsection{Stage 3: board warp perturbation}

When further model expansion is required, a radiometric warp candidate is evaluated without requiring the geometry model to be retained. When global geometry is also supported by the data, a geometry--warp candidate is considered. Among statistically adequate candidates, the model with the lower Bayesian Information Criterion (BIC) is retained. Consequently, PCB warp does not automatically inherit unnecessary global-geometry DoF.

For large plate-like PCBs, local reference layers may move axially due to the bow or twist of the board. To provide a controlled low-dimensional stress test for board deformation, we used a representative quadratic bow–twist field
\begin{equation}
w(x,y)=
a_w
\left[
a_b (X_n^2 + Y_n^2 - c)
+
a_t X_n Y_n
\right],
\label{eq:warp}
\end{equation}
where $X_n$ and $Y_n$ are normalized in-plane coordinates, $c$ is a constant chosen to remove a piston offset, and $a_b$ and $a_t$ are the weight coefficients of the bow and twist/saddle, respectively. The quadratic radial term represents smooth bow-like curvature, while the cross term introduces a twist/saddle component, consistent with common PCB bow/twist descriptions and second-order polynomial representations of electronic package warpage \cite{hubble2017surface}. The coefficients define the particular mixed deformation used in this stress test and are not intended to represent a universal PCB warpage law.

\subsection{Statistical adequacy gate}

After convergence of a stage, we calculate a weighted statistic of the measured region after removing per-view log offsets:
\begin{equation}
\tilde{r}_j = r_j - b_{v(j)}, \quad
\chi^2 = \sum_{j\in\Omega_T} W_j \tilde{r}_j^2,
\end{equation}
where $\Omega_T$ contains task-sensitive pixels, $r_j$ is the residual at the detector pixel $j$, $b_{v(j)}$ is the removed offset. With effective DoF $N$,
\begin{equation}
\chi_r^2=\chi^2/{N}.
\end{equation}
The model is considered adequate when
\begin{equation}
\chi_r^2\le
T_\chi,\qquad
T_\chi=
\max\left[
1.35,\,
1+4\sqrt{2/N}
\right].
\label{eq:gate}
\end{equation}
The fixed floor tolerates a modest approximation in the log-noise model, while the statistical term shrinks with increasing sample count. If the current stage fails Eq.~(\ref{eq:gate}), a more complex nested model is evaluated \cite{kass1995bayes}. It is accepted only if
\begin{equation}
\Delta\mathrm{BIC}
=
\mathrm{BIC}_{\mathrm{simple}}
-
\mathrm{BIC}_{\mathrm{complex}}
\ge 10,
\label{eq:bicgate}
\end{equation}
\begin{equation}
\mathrm{BIC}=\chi^2+k\log{n_D},
\end{equation}
where $k$ is the number of free parameters estimated in the candidate model, $n_D$ the number of data samples used. This threshold enforces strong evidence before adding perturbation DoF.

\section{Task information under perturbation uncertainty}

For a parameter vector in Eq.~(\ref{eq:para}), the log-domain Fisher approximation is $F$=$H_{\mathrm{data}}$, which can be partitioned as
\begin{equation}
 F=
\begin{bmatrix}
F_{LL} &  F_{L\eta}\\
 F_{\eta L} &  F_{\eta\eta}
\end{bmatrix},
\end{equation}
the known-perturbation or conditional Cramér–Rao lower bound (CRLB) is $\sigma_{L,\mathrm{cond}} =1/\sqrt{F_{LL}}$. When perturbation parameters must be estimated jointly, the effective task information is the Schur complement
\begin{equation}
\mathcal I_{L,\mathrm{eff}}
=
F_{LL}
-
 F_{L\eta}
 F_{\eta\eta}^{-1}
 F_{\eta L}, \quad 
 \sigma_{L,\mathrm{marg}} = 1/\sqrt{\mathcal I_{L,\mathrm{eff}}}.
\label{eq:schur}
\end{equation}

If the calibration prior $F_{\mathrm{prior}}$ is available,
$F_{\mathrm{post}} = F_{\mathrm{data}} + F_{\mathrm{prior}}$ is used to obtain a prior-aware hybrid bound. If the data-only perturbation block is numerically singular, the corresponding data-only marginalized bound is reported as infinite rather than using a pseudoinverse that could imply spurious identifiability. Because the estimator is hierarchical, we calculate stage-specific bounds for radiometric, geometric, and warp perturbation families. For matched-model view and photon sweeps, the radiometric-stage marginalized bound is the relevant reference because the adequacy gate should prevent activation of higher-complexity models.

\section{Experimental setup}

\subsection{Canonical layered phantom}

A parametric layered phantom was constructed to isolate axial feature information while retaining PCB-like physics (Fig.~\ref{fig:planarCT}(b)). Rather than using a fixed voxelized volume, the phantom is defined by adjustable parameters including lateral ROI, board thickness, axial termination gap \(L\), reference-plane position \(Z_{\rm ref}\), Cu reference-layer thickness and Cu/FR4 material assignment. It also contains multiple Cu reference layers, asymmetric Cu context features, and a cylindrical via with a terminating axial edge. The nominal configuration used a \(4.8\times4.8~\mathrm{mm}^2\) ROI, a \(3.2~\mathrm{mm}\) board thickness, \(L=120~\mu\mathrm{m}\), \(Z_{\rm ref}=0.55~\mathrm{mm}\), and a \(35~\mu\mathrm{m}\) Cu reference layer (see Supplementary Materials, Table~S1).

The termination voxel is modeled with continuous axial occupancy so that subvoxel changes in \(L\) produce continuous projection changes, allowing meaningful finite-difference sensitivities below the reconstruction voxel spacing. The estimator is therefore tied to the assumed low-dimensional structural model rather than to one fixed phantom realization: represented geometric parameters can be varied without modifying the inference framework, whereas unmodeled structural variations constitute model mismatch.

\subsection{Acquisition and reconstruction}

The nominal geometry used SOD = $120~\mathrm{mm}$, SDD = $360~\mathrm{mm}$, and obliquity $35^\circ$. The source used a polychromatic spectrum with nominal tube potential $120~\mathrm{kVp}$; the detector model includes a finite point spread function (PSF), focal-spot unsharpness, Poisson statistics, read noise, gain variation, and finite flat-field averaging. Detector dimensions guarantee complete projected-volume coverage. Reconstruction baselines included FDK\cite{feldkamp1984practical}, CGLS\cite{hestenes1952methods}, SIRT\cite{gilbert1972iterative,lakshminarayanan1979methods}, and SIRT followed by 3D total variation denoising (SIRT-TV)\cite{rudin1992nonlinear}. The reported results use the SIRT-TV reconstruction-driven metrology baseline. All GPU forward projections and reconstructions were implemented with ASTRA CUDA \cite{vanAarle2016fast}. Geometry conventions were separately validated by cone-vector parity tests and 3D marker localization prior to statistical experiments. The details of the parameters are summarized in the Supplementary Materials (Table~S2).

\subsection{View-number and photon-budget sweeps}

The view sweep used $N_v\in\{12,18,24,32,48,64\}$ under a fixed total incident photon budget:
\begin{equation}
N_v N_{\gamma}^{(\mathrm{view})}
=
N_{\gamma,\mathrm{tot}}.
\end{equation}
The photon sweep varied incident photons per detector pixel from $2\times10^4$ to $10^6$ at fixed geometry.

\subsection{Forward-model perturbation}

\begin{figure}[!h]
\centering
\includegraphics[width=\linewidth]{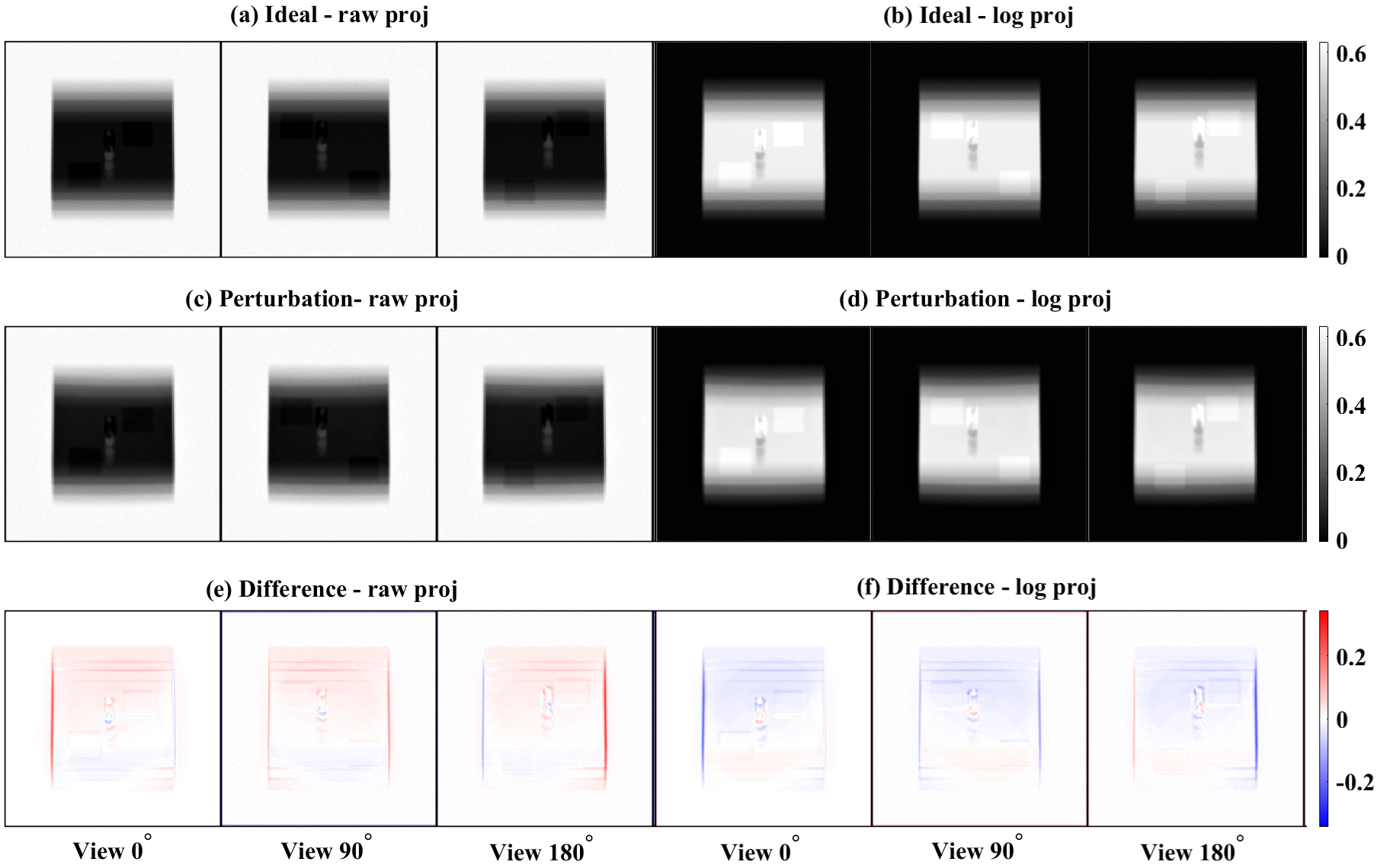}
\caption{Examples of projection images of the ideal forward model and another model with combined perturbations. (a,c) the raw projection images for the two models. (b,d) the log projection images for the two models. (e,f) the difference between the two models. Color bars are only for the log projections.}
\label{fig:projection}
\end{figure}

The hierarchical estimator was evaluated under matched data and seven perturbation conditions. The spectrum perturbation changed tube voltage and filtration; the blur perturbation changed detector PSF and focal spot; the scatter perturbation introduced a $5\%$ low-frequency component. The global geometric perturbation changed magnification, obliquity, and detector-$v$ alignment. The board warp used Eq.~(\ref{eq:warp}) with $a_w=120~\mu\mathrm{m}$. An additional geometry-jitter case applied independent per-view source/detector perturbations, intentionally outside the low-dimensional perturbation family. The combined case included spectrum, blur, scatter, global geometry, board warp, and residual random per-view jitter. Fig.~\ref{fig:projection} illustrates the images from three projection views of the ideal forward model and another one with combined perturbations.

\subsection{Statistical analysis}

Each setting of matched model, view number, photon budget and perturbation were evaluated using 30 independent Monte Carlo trials. Bias, standard deviation (SD), and root mean square error (RMSE) were calculated across trials. The confidence intervals (CIs) for the error statistics were obtained using 10,000 nonparametric bootstrap resamples. Stage-activation and adequacy-pass rates were calculated as empirical proportions across trials.

\section{Results}

The results are organized around three computational sensing questions: whether the low-dimensional axial metrology can be estimated with fewer measurements than are required for image reconstruction, whether the observed precision follows the Fisher limit of the task, and whether physically modeled perturbations can be distinguished from unmodeled acquisition errors.

\subsection{Matched-model performance after unbiased transmission preprocessing}

With the unbiased transmission preprocessing of Eq.~(5), the matched-model experiment was carried out. The direct projection-domain estimator exhibited a bias of only $-0.018$ $\mu m$, an empirical SD of 0.152 $\mu m$, and an RMSE of 0.150 $\mu m$ (95\% bootstrap CIs: 0.118–0.179 $\mu m$). The reduced $\chi^2$ remained essentially unity, $\chi_r^2=1.000$, indicating statistical consistency between the measurements and the assumed physical model. The adequacy gate passed in all realizations, while neither the geometric nor the warp perturbation model was activated. Therefore, after the removal of preprocessing-induced bias, the task-driven direct estimator is effectively unbiased at the submicrometer scale and that the hierarchical model does not introduce unnecessary perturbation DoF when the nominal forward model is adequate.

\subsection{Task-information saturation with view number}

\begin{figure}[!h]
\centering
\includegraphics[width=0.75\linewidth]{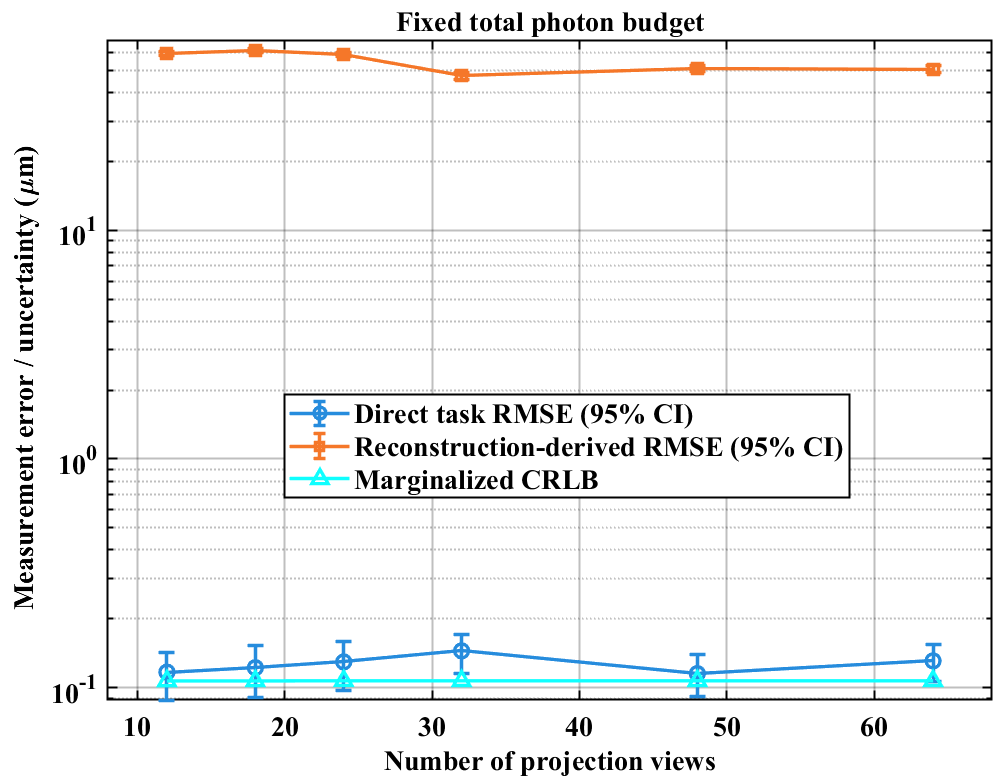}
\caption{Axial metrology performance versus projection-view number under a fixed total photon budget. Direct projection-domain measurement RMSE remains close to the radiometric-stage marginalized CRLB across 12--64 views, whereas reconstruction-driven measurement RMSE remains tens of micrometers and strongly view-dependent. Error bars denote 95\% CIs.}
\label{fig:viewtask}
\end{figure}

As shown in Fig.~\ref{fig:viewtask}, increasing the number of views from 12 to 64 under a fixed total photon budget produced little change in the direct projection-domain estimate. In all six settings, the bias remained between approximately $-0.054$ and $0.021~\mu\mathrm{m}$, the RMSE between $0.115$ and $0.145~\mu\mathrm{m}$, and the bootstrap 95\% CI overlapped substantially. The marginalized CRLB was nearly constant at approximately $0.107~\mu\mathrm{m}$, indicating that the ATG-relevant information was already close to saturation at the lowest view count tested. In contrast, reconstruction-driven metrology remained strongly sampling-dependent, with RMSEs of $59.2~\mu\mathrm{m}$ at 12 views, $61.0~\mu\mathrm{m}$ at 18 views, $47.3~\mu\mathrm{m}$ at 32 views and approximately $51~\mu\mathrm{m}$ at 48--64 views.

As such, the angular sampling requirement for task-driven ATG estimation in the projection domain is substantially lower than that for reconstruction-driven metrology in the image domain. Once the sensitivity subspace associated with $L$ is sufficiently sampled, additional views provide little additional task information under a fixed photon budget, even though they may still benefit the intermediate 3D reconstruction.

\subsection{Photon-budget scaling}

\begin{figure}[!h]
\centering
\includegraphics[width=0.75\linewidth]{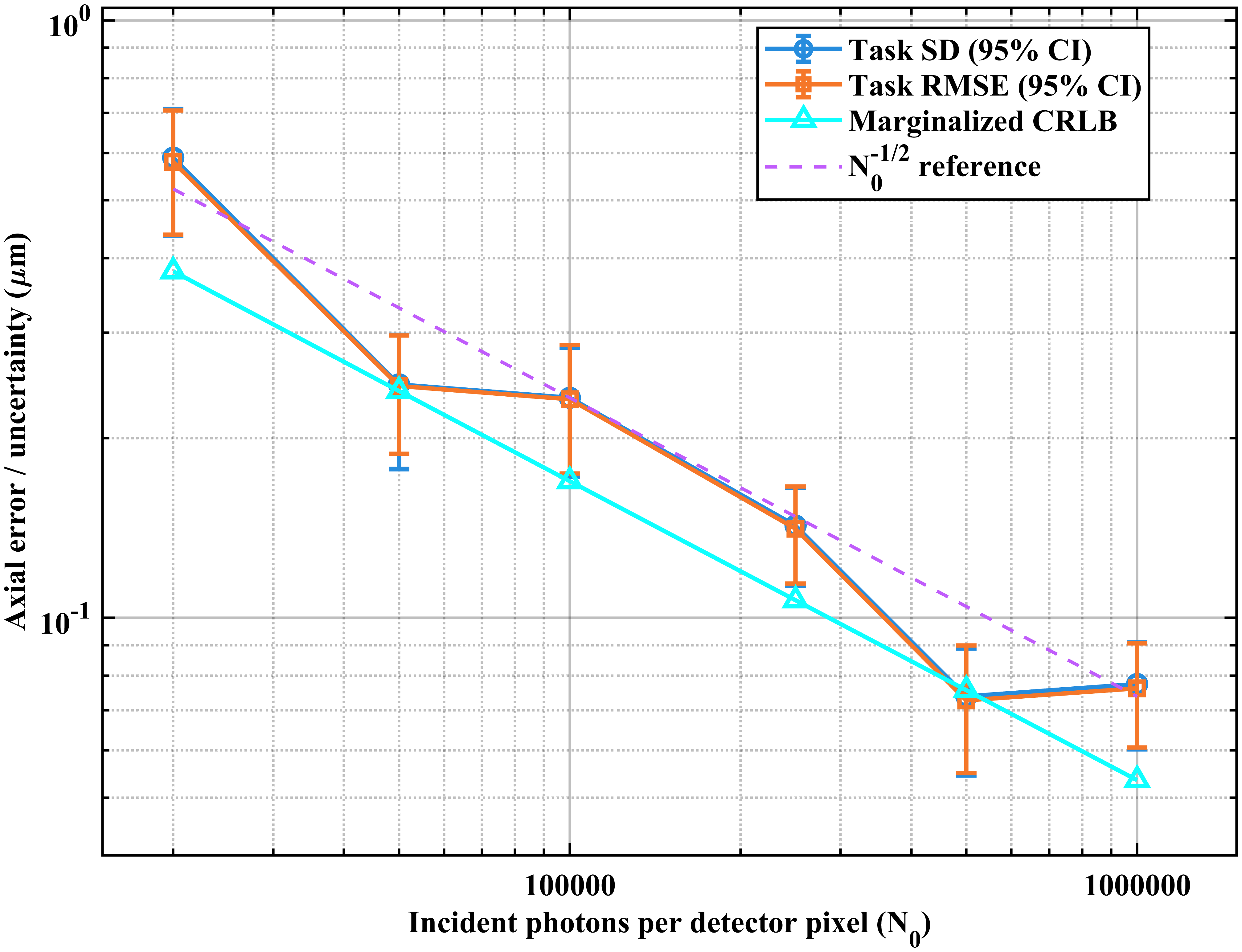}
\caption{Projection-domain measurement uncertainty versus incident photons per pixel. The SD and the RMSE track the Fisher-information trend. Error bars denote 95\% CIs.}
\label{fig:photon}
\end{figure}

Fig.~\ref{fig:photon} compares the empirical projection-domain uncertainty with the marginalized CRLB of the radiometric-stage over the photon-budget range tested. Both the SD and the RMSE decrease systematically as the incident photon count increases, closely following the expected photon-limited inverse square-root scaling. A log--log fit gives slopes of $-0.524$ for the empirical SD and $-0.525$ for the RMSE, while the CRLB exhibits a slope of $-0.501$, in agreement with the theoretical dependence $N_0^{-1/2}$. The close agreement of these slopes indicates that the direct ATG estimator remains predominantly photon-noise limited over the investigated photon-budget range and preserves the expected scaling of information with photon count. The empirical uncertainty remains of the same order as the predicted Fisher limit, while deviations at individual photon-budget levels are attributable to finite Monte Carlo sampling and residual approximations in the noise and perturbation models. Importantly, no systematic increase in estimation bias is observed as the photon budget is reduced, supporting the hypothesis that the observed noise-dependent bias is associated with transmission preprocessing rather than with the estimator itself.

\subsection{Hierarchical correction of modeled perturbations}

Table~\ref{tab:mismatch} summarizes the gated hierarchical estimator under matched and perturbed conditions. For the matched-model, spectrum, blur and scatter cases, the estimator remained in the radiometric stage, the reduced $\chi^2$ remained near unity, and the adequacy gate passed in all 30 trials. The corresponding projection-domain RMSEs were $0.150$, $0.112$, $0.195$, and $0.144~\mu\mathrm{m}$, respectively. These modeled radiometric perturbations were handled without activating unnecessary geometric degrees of freedom.

The global-geometry case provides a quantitative stress test of systematic alignment errors. With imposed perturbations of approximately $-5.81\times10^{-3}$ in relative magnification, $+0.30^\circ$ in obliquity, and $+0.12~\mathrm{mm}$ in the detector-$v$ position, the geometry stage was activated in all realizations. The ATG RMSE remained $0.153~\mu\mathrm{m}$ and the mean reduced $\chi^2$ was $1.012$. The recovered values, $-5.81\times10^{-3}$, $0.3000^\circ$, and $0.1198~\mathrm{mm}$, closely matched the imposed perturbations. Therefore, representing global geometry through relative magnification captures the dominant alignment effect while avoiding the strong degeneracy of separately estimating SOD and SDD.

For the board warp, the warp stage was activated in all trials, while the geometry model was activated only in $13.3\%$ of the cases. The recovered warp amplitude was $119.97~\mu\mathrm{m}$ for an imposed $120~\mu\mathrm{m}$, with an ATG RMSE of $0.136~\mu\mathrm{m}$. These results show that the hierarchy can distinguish specimen deformation from global acquisition-geometry errors while retaining submicrometer task accuracy.

\begin{table*}[!t]
\centering
\caption{Hierarchical physical perturbation results. (Bias, SD, RMSE and CI are in $\mu$m. GEO: geometry; GA: geometry activated; WA: warp activated).}
\setlength{\tabcolsep}{3pt}
\label{tab:mismatch}
\begin{tabular}{lccccccccc}
\toprule
Condition & Bias & SD & RMSE & RMSE CI & $\overline{\chi_r^2}$ & Pass rate & GA rate & WA rate \\
\midrule
Matched & -0.0175 & 0.1518 & 0.1503 & 0.1181--0.1791 & 1.0000 & 1.0000 & 0.0000 & 0.0000 \\
Spectrum & -0.0195 & 0.1118 & 0.1117 & 0.0920--0.1305 & 1.0003 & 1.0000 & 0.0000 & 0.0000 \\
Blur & -0.0271 & 0.1960 & 0.1946 & 0.1474--0.2386 & 1.0000 & 1.0000 & 0.0000 & 0.0000 \\
Scatter & 0.0028 & 0.1467 & 0.1443 & 0.1024--0.1843 & 0.9998 & 1.0000 & 0.0000 & 0.0000 \\
GEO global & 0.1034 & 0.1148 & 0.1530 & 0.1212--0.1819 & 1.0119 & 1.0000 & 1.0000 & 0.0000 \\
Board warp & 0.0239 & 0.1364 & 0.1362 & 0.1034--0.1679 & 0.9999 & 1.0000 & 0.1333 & 1.0000 \\
Combined & -0.8509 & 0.6437 & 1.0604 & 0.8819--1.2189 & 5.1231 & 0.0000 & 1.0000 & 1.0000 \\
GEO jitter & -5.5469 & 4.4051 & 7.0374 & 5.7732--8.4538 & 20.0674 & 0.0000 & 0.1667 & 0.0000 \\
\bottomrule
\end{tabular}

\end{table*}

\begin{figure}[!ht]
\centering
\includegraphics[width=0.8\linewidth]{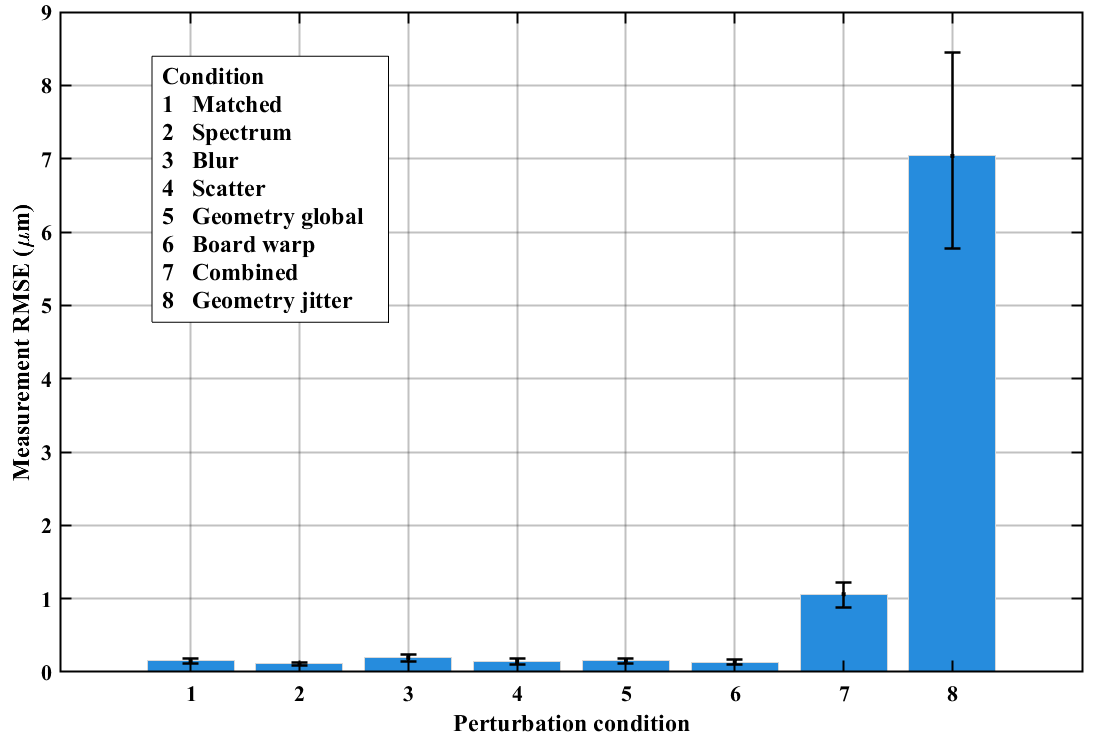}
\caption{Projection-domain ATG RMSE under matched and perturbed forward models. Modeled radiometric, global-geometry, and board-warp perturbations retain sub-micrometer task error, whereas the combined and random per-view geometry-jitter conditions contain residual components outside the fitted low-dimensional perturbation family. Error bars denote 95\% CIs.}
\label{fig:mismatch}
\end{figure}

\subsection{Detection of residual and out-of-model perturbations}

The combined perturbation activated both geometry and warp correction in all cases. Although the dominant perturbation parameters were recovered close to their imposed values, the residual model mismatch remained: the projection-domain RMSE increased to $1.06~\mu\mathrm{m}$ with a bias of $-0.851~\mu\mathrm{m}$, while the reduced $\chi^2$ remained elevated at $5.12$ and the adequacy gate failed in all cases.

\begin{figure}[!h]
\centering
\includegraphics[width=0.8\linewidth]{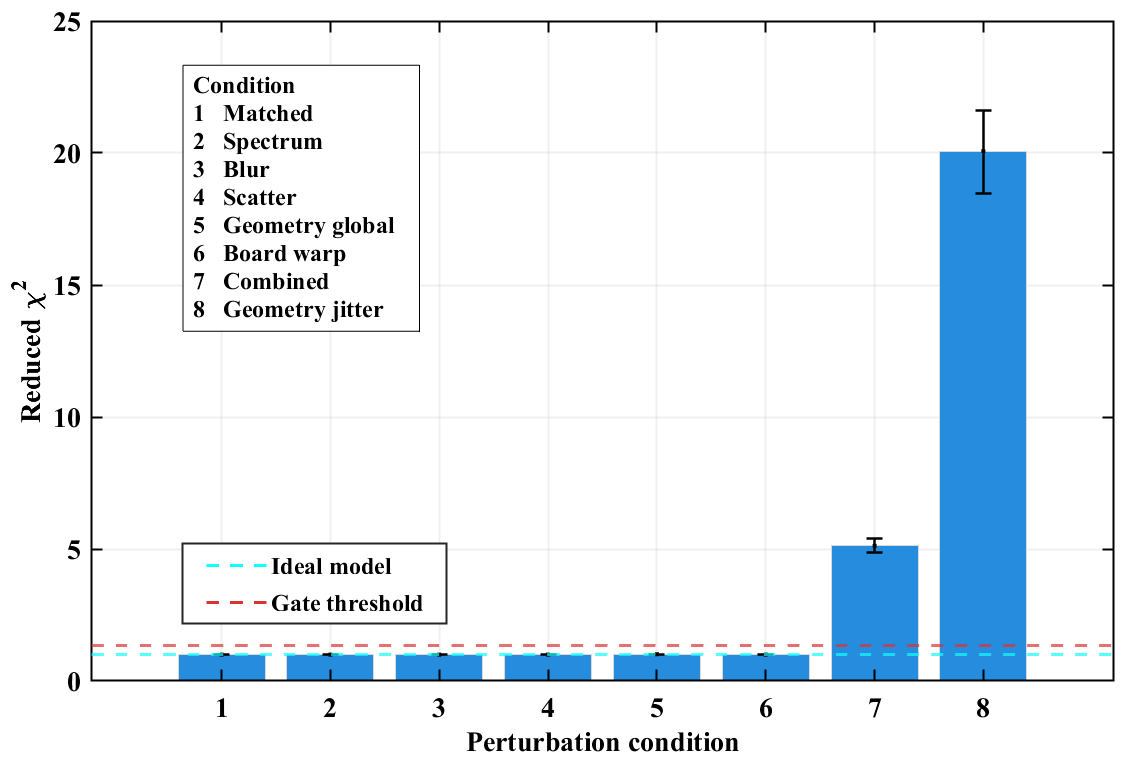}
\caption{Reduced-$\chi^2$ model-adequacy statistic after estimation. Matched and successfully modeled perturbations return values near the expected noise level, whereas combined perturbations and random geometry jitter remain strongly elevated. The dashed line denotes the fixed adequacy threshold. Error bars denote 95\% CIs.}
\label{fig:adequacy}
\end{figure}

A stronger failure occurred for random per-view geometry jitter, which is intentionally outside the assumed global-geometry model. Its RMSE increased to $7.04~\mu\mathrm{m}$ and the mean reduced $\chi^2$ to $20.1$; the adequacy gate failed in all 30 trials, geometry was activated in only $16.7\%$ of the cases and board warp was never activated. Thus, rather than absorbing unmodeled view-dependent motion into unrelated perturbation parameters, the hierarchy correctly identifies the model inadequacy. As shown in Fig.~\ref{fig:adequacy}, successfully modeled conditions return to $\chi_r^2\simeq1$, while the combined and per-view jitter cases remain elevated due to residual components outside the fitted low-dimensional perturbation family.

This behavior is useful for metrology. The convergence of an optimizer or a small apparent random variance does not guarantee physical accuracy if the forward model is wrong. A separate model-adequacy statistic provides a principled mechanism for flagging cases in which the reported estimate should not be trusted without recalibration or a richer physical model.

\section{Discussion}

\subsection{Task-driven and reconstruction-driven metrology have different sampling requirements}

The fixed-photon-budget experiment shows that task-driven metrology can require substantially fewer sampling views than the traditional reconstruction-driven approach. Here, ATG estimation depends on the projection-domain sensitivity to a single axial parameter after marginalizing active perturbation directions, whereas 3D reconstruction seeks to recover a much higher-dimensional attenuation distribution. The two inverse problems therefore do not need to share the same sampling requirement.

The nearly invariant marginalized CRLB over 12--64 views indicates that most of the ATG-relevant information is already captured at the lower end of the tested range. Redistributing the same total photon budget over additional views increases angular sampling density but contributes little additional Fisher information for the measurement \cite{ly2017tutorial}, while reconstruction-driven ATG estimation remains more view-dependent because localization of the reference layer and feature termination require a richer spatial representation.

This does not imply that 12 views are universally sufficient or optimal; the saturation point depends on the task, object structure, acquisition geometry, noise level, and model uncertainty. Rather, acquisition sufficiency for a predefined quantitative task should be assessed from task error and marginalized Fisher information rather than inherited from the requirements of general-purpose 3D reconstruction. Once the task-relevant measurement subspace is adequately sampled, additional projections may offer diminishing returns for metrology, even though they remain useful for other imaging objectives.

\subsection{Role of reconstruction in task-driven XCT}

The proposed formulation does not eliminate the need for 3D reconstruction, which remains valuable for visualization, defect inspection, and multi-task analysis. Its main advantage arises when the measurement objective is known in advance, allowing reconstruction to serve as an auxiliary representation rather than a mandatory intermediate step. Consequently, conventional image-quality metrics such as PSNR and SSIM \cite{brunet2011mathematical} may not reliably predict metrology performance: visibly degraded reconstructions can preserve task-relevant projection signatures, whereas visually plausible images may contain systematic axial shifts that bias the measurement. Therefore, task-driven acquisition should be evaluated using task error, Fisher information, and model adequacy in addition to conventional image metrics.

\subsection{Complexity control and model adequacy}

The perturbation experiments highlight two roles of the hierarchical formulation. First, model complexity should increase only when necessary, because perturbation directions that are strongly correlated with the task sensitivity can absorb task-relevant signal and reduce identifiability. The observed stage selection therefore favors statistical gating over unrestricted simultaneous fitting of all perturbations.

Second, numerical convergence does not guarantee physical model adequacy. When the assumed perturbation family explains the data, the normalized residual returns to the expected noise scale; when important errors remain outside that family, the estimator may still produce a value of $\widehat L$, but an elevated reduced-$\chi^2$ indicates that the estimate is not supported by the assumed measurement model \cite{possolo2015simple}. In practice, failure of the adequacy test could trigger recalibration, additional views, or a more general reconstruction procedure rather than reporting a potentially biased measurement.

\subsection{Precision, accuracy, and current limitations}

The reported sub-micrometer errors characterize statistical precision and modeled-perturbation performance within the simulated system; they do not imply sub-micrometer absolute accuracy for an uncalibrated instrument. Absolute accuracy also depends on spectrum and detector calibration, focal-spot characterization, geometry, material attenuation, scatter, mechanical stability, and specimen deformation.

The physical model is intentionally low-dimensional: the spectrum and attenuation are approximate, the scatter is represented without full particle transport, and the warp model captures only a restricted deformation family. The random per-view jitter remains outside the estimator model \cite{chetley2011estimation}. These limitations define the scope of future experimental validation rather than requiring further expansion of the present estimator. Physical experiments should therefore calibrate the X-ray spectrum and geometry and compare projection-domain metrology with an external reference. Similarly, current alignment studies should be interpreted as representative stress tests rather than a complete tolerance map, which would require instrument-specific mechanical and calibration specifications.

\subsection{Generality of the computational sensing formulation}

Although PCB backdrill metrology is used as a test case, the formulation is broader. Any XCT problem in which the desired output can be represented by low-dimensional parameterization is a candidate for task-driven direct inference. Examples include interlayer spacing, local bond-line thickness, buried-interface position, void depth, plated-wall thickness, and axial localization of defects in advanced packaging \cite{chai2026tsv}. The same concept also applies beyond XCT, such as Fourier ptychography \cite{konda2020fourier, zheng2021concept, li2025fourier}: when a physical forward model maps a small set of task parameters into a much higher-dimensional measurement, acquisition can be optimized around the corresponding sensitivity subspace rather than around complete image reconstruction.

\section{Conclusion}

We present a task-driven computational X-ray sensing framework for sparse-view axial metrology that estimates the target parameter directly from projection data rather than from a reconstructed 3D volume. A gated low-dimensional perturbation model accounts for radiometric, geometric, and specimen-deformation effects only when necessary, while a perturbation-marginalized Fisher analysis quantifies task-specific information and uncertainty. Numerical results show that, under a fixed total photon budget, direct metrology is nearly insensitive to view count over 12–64 projections, whereas reconstruction-driven metrology remains strongly sampling-dependent with errors in the tens of micrometers. The direct estimator follows the expected photon-limited scaling, maintains sub-micrometer accuracy for modeled perturbations, and flags out-of-model errors through an elevated model-adequacy statistic. These results support designing XCT acquisition and inference around the information required by the quantitative task rather than the fidelity requirements of general-purpose 3D reconstruction.\\

\textbf{Funding}\\
This work was partially supported by Shenzhen Talent Startup Funds (No. 827-000954).

\textbf{Disclosures}\\
\noindent YT: Litemaze Technology (F,I).

\textbf{Data availability} \\
Parameters for the experiments are in the Appendix. Other data underlying the results presented in this paper may be obtained from the authors upon reasonable request.


\bibliographystyle{unsrtnat}
\bibliography{references}  

@book{kak1988principles,
  author    = {Kak, Avinash C. and Slaney, Malcolm},
  title     = {Principles of Computerized Tomographic Imaging},
  publisher = {IEEE Press},
  year      = {1988}
}

@article{feldkamp1984practical,
  author  = {Feldkamp, L. A. and Davis, L. C. and Kress, J. W.},
  title   = {Practical cone-beam algorithm},
  journal = {J. Opt. Soc. Am. A},
  volume  = {1},
  number  = {6},
  pages   = {612--619},
  year    = {1984}
}

@article{andersen1984simultaneous,
  author  = {Andersen, A. H. and Kak, A. C.},
  title   = {Simultaneous algebraic reconstruction technique ({SART}): a superior implementation of the {ART} algorithm},
  journal = {Ultrasonic Imaging},
  volume  = {6},
  number  = {1},
  pages   = {81--94},
  year    = {1984}
}

@article{sidky2008image,
  author  = {Sidky, Emil Y. and Pan, Xiaochuan},
  title   = {Image reconstruction in circular cone-beam computed tomography by constrained, total-variation minimization},
  journal = {Phys. Med. Biol.},
  volume  = {53},
  number  = {17},
  pages   = {4777--4807},
  year    = {2008}
}

@article{tuy1983inversion,
  author  = {Tuy, Heang K.},
  title   = {An inversion formula for cone-beam reconstruction},
  journal = {SIAM J. Appl. Math.},
  volume  = {43},
  number  = {3},
  pages   = {546--552},
  year    = {1983}
}

@article{helfen2005high,
  author  = {Helfen, L. and Baumbach, T. and Mikul{\'i}k, P. and Kiel, D. and Pernot, P. and Cloetens, P. and Baruchel, J.},
  title   = {High-resolution three-dimensional imaging of flat objects by synchrotron-radiation computed laminography},
  journal = {Appl. Phys. Lett.},
  volume  = {86},
  pages   = {071915},
  year    = {2005}
}

@article{vanAarle2015astra,
  author  = {van Aarle, Wim and Palenstijn, Willem Jan and De Beenhouwer, Jan and Altantzis, Thomas and Bals, Sara and Batenburg, K. Joost and Sijbers, Jan},
  title   = {The {ASTRA} Toolbox: A platform for advanced algorithm development in electron tomography},
  journal = {Ultramicroscopy},
  volume  = {157},
  pages   = {35--47},
  year    = {2015}
}

@article{vanAarle2016fast,
  author  = {van Aarle, Wim and Palenstijn, Willem Jan and Cant, Jeroen and Janssens, Eline and Bleichrodt, Folkert and Dabravolski, Andrei and De Beenhouwer, Jan and Batenburg, K. Joost and Sijbers, Jan},
  title   = {Fast and flexible {X}-ray tomography using the {ASTRA} toolbox},
  journal = {Opt. Express},
  volume  = {24},
  number  = {22},
  pages   = {25129--25147},
  year    = {2016}
}

@book{barrett2004foundations,
  author    = {Barrett, Harrison H. and Myers, Kyle J.},
  title     = {Foundations of Image Science},
  publisher = {Wiley},
  year      = {2004}
}

@book{kay1993fundamentals,
  author    = {Kay, Steven M.},
  title     = {Fundamentals of Statistical Signal Processing, Volume I: Estimation Theory},
  publisher = {Prentice Hall},
  year      = {1993}
}

@article{schwarz1978estimating,
  author  = {Schwarz, Gideon},
  title   = {Estimating the dimension of a model},
  journal = {Ann. Statist.},
  volume  = {6},
  number  = {2},
  pages   = {461--464},
  year    = {1978}
}

@article{lin2023deep,
  author  = {Lin, Yiqun and Luo, Zhongjin and Zhao, Wei and Li, Xiaomeng},
  title   = {Learning deep intensity field for extremely sparse-view {CBCT} reconstruction},
  journal = {arXiv preprint arXiv:2303.06681},
  year    = {2023}
}

@article{barbastathis2024sensitivity,
  author  = {Barbastathis, George},
  title   = {Sensitivity fields and parameter estimation from dielectric objects},
  journal = {J. Opt. Soc. Am. A},
  volume  = {41},
  number  = {11},
  pages   = {C82--C93},
  year    = {2024}
}

@article{mait2018computational,
  title={Computational imaging},
  author={Mait, Joseph N and Euliss, Gary W and Athale, Ravindra A},
  journal={Advances in Optics and Photonics},
  volume={10},
  number={2},
  pages={409--483},
  year={2018},
  publisher={Optical Society of America}
}

@article{sitzmann2018end,
  title={End-to-end optimization of optics and image processing for achromatic extended depth of field and super-resolution imaging},
  author={Sitzmann, Vincent and Diamond, Steven and Peng, Yifan and Dun, Xiong and Boyd, Stephen and Heidrich, Wolfgang and Heide, Felix and Wetzstein, Gordon},
  journal={ACM Transactions on Graphics},
  volume={37},
  number={4},
  pages={1--13},
  year={2018},
  publisher={ACM New York, NY, USA}
}

@article{wang2022differentiable,
  title={dO: A differentiable engine for deep lens design of computational imaging systems},
  author={Wang, Congli and Chen, Ni and Heidrich, Wolfgang},
  journal={IEEE Transactions on Computational Imaging},
  volume={8},
  pages={905--916},
  year={2022},
  publisher={IEEE}
}

@article{arguello2023deep,
  title={Deep optical coding design in computational imaging: a data-driven framework},
  author={Arguello, Henry and Bacca, Jorge and Kariyawasam, Hasindu and Vargas, Edwin and Marquez, Miguel and Hettiarachchi, Ramith and Garcia, Hans and Herath, Kithmini and Haputhanthri, Udith and Ahluwalia, Balpreet Singh and others},
  journal={IEEE Signal Processing Magazine},
  volume={40},
  number={2},
  pages={75--88},
  year={2023},
  publisher={IEEE}
}

@article{islam2023bending,
  title={Bending analysis of glass fiber reinforced epoxy composites/copper-clad laminates for multi-layer printed circuit boards},
  author={Islam, Md Nazmul and Anwar, Md Sayed and Islam, Md Shariful and Arifuzzaman, Md and Al Bari, Md Abdullah},
  journal={Hybrid Advances},
  volume={4},
  pages={100090},
  year={2023},
  publisher={Elsevier}
}

@article{alvarez1976energy,
  title={Energy-selective reconstructions in x-ray computerised tomography},
  author={Alvarez, Robert E and Macovski, Albert},
  journal={Physics in Medicine \& Biology},
  volume={21},
  number={5},
  pages={733--744},
  year={1976}
}

@article{yang2020x,
  title={X-ray attenuation models to account for beam hardening in computed tomography},
  author={Yang, Qiheng and Fullagar, WK and Myers, GR and Latham, SJ and Varslot, Trond and Sheppard, AP and Kingston, AM},
  journal={Applied Optics},
  volume={59},
  number={29},
  pages={9126--9136},
  year={2020},
  publisher={Optical Society of America}
}

@article{tilley2016model,
  title={Model-based iterative reconstruction for flat-panel cone-beam {CT} with focal spot blur, detector blur, and correlated noise},
  author={Tilley, Steven and Siewerdsen, Jeffrey H and Stayman, J Webster},
  journal={Physics in Medicine \& Biology},
  volume={61},
  number={1},
  pages={296--319},
  year={2016},
  publisher={IOP Publishing}
}

@article{mccann2021local,
  title={Local models for scatter estimation and descattering in polyenergetic X-ray tomography},
  author={McCann, Michael T and Klasky, Marc L and Schei, Jennifer L and Ravishankar, Saiprasad},
  journal={Optics Express},
  volume={29},
  number={18},
  pages={29423--29438},
  year={2021},
  publisher={Optical Society of America}
}

@article{xu2009electronic,
  title={Electronic noise modeling in statistical iterative reconstruction},
  author={Xu, Jingyan and Tsui, Benjamin MW},
  journal={IEEE Transactions on Image Processing},
  volume={18},
  number={6},
  pages={1228--1238},
  year={2009},
  publisher={IEEE}
}

@article{ding2018statistical,
  title={Statistical image reconstruction using mixed Poisson-Gaussian noise model for X-ray {CT}},
  author={Ding, Qiaoqiao and Long, Yong and Zhang, Xiaoqun and Fessler, Jeffrey A},
  journal={arXiv preprint arXiv:1801.09533},
  year={2018}
}

@article{van2015dynamic,
  title={Dynamic intensity normalization using eigen flat fields in X-ray imaging},
  author={Van Nieuwenhove, Vincent and De Beenhouwer, Jan and De Carlo, Francesco and Mancini, Lucia and Marone, Federica and Sijbers, Jan},
  journal={Optics Express},
  volume={23},
  number={21},
  pages={27975--27989},
  year={2015},
  publisher={Optical Society of America}
}

@article{fu2016comparison,
  title={Comparison between pre-log and post-log statistical models in ultra-low-dose {CT} reconstruction},
  author={Fu, Lin and Lee, Tzu-Cheng and Kim, Soo Mee and Alessio, Adam M and Kinahan, Paul E and Chang, Zhiqian and Sauer, Ken and Kalra, Mannudeep K and De Man, Bruno},
  journal={IEEE Transactions on Medical Imaging},
  volume={36},
  number={3},
  pages={707--720},
  year={2017},
  publisher={IEEE}
}

@article{hestenes1952methods,
  title={Methods of conjugate gradients for solving linear systems},
  author={Hestenes, Magnus R and Stiefel, Eduard and others},
  journal={Journal of Research of the National Bureau of Standards},
  volume={49},
  number={6},
  pages={409--436},
  year={1952}
}

@article{gilbert1972iterative,
  title={Iterative methods for the three-dimensional reconstruction of an object from projections},
  author={Gilbert, Peter},
  journal={Journal of Theoretical Biology},
  volume={36},
  number={1},
  pages={105--117},
  year={1972},
  publisher={Elsevier}
}

@article{lakshminarayanan1979methods,
  title={Methods of least squares and SIRT in reconstruction},
  author={Lakshminarayanan, AV and Lent, A},
  journal={Journal of Theoretical Biology},
  volume={76},
  number={3},
  pages={267--295},
  year={1979},
  publisher={Elsevier}
}

@article{rudin1992nonlinear,
  title={Nonlinear total variation based noise removal algorithms},
  author={Rudin, Leonid I and Osher, Stanley and Fatemi, Emad},
  journal={Physica D: Nonlinear Phenomena},
  volume={60},
  number={1-4},
  pages={259--268},
  year={1992},
  publisher={Elsevier}
}

@article{hubble2017surface,
  title={Surface Mount Signed Warpage Case Study; New Methods for Characterizing 3D Shapes Through Reflow Temperatures},
  author={Hubble, Neil and Young, Jerry and Hartnett, Kim},
  journal={Proceedings of IPC APEX},
  year={2017}
}

@article{brunet2011mathematical,
  title={On the mathematical properties of the structural similarity index},
  author={Brunet, Dominique and Vrscay, Edward R and Wang, Zhou},
  journal={IEEE Transactions on Image Processing},
  volume={21},
  number={4},
  pages={1488--1499},
  year={2011},
  publisher={IEEE}
}

@article{kass1995bayes,
  title={Bayes factors},
  author={Kass, Robert E and Raftery, Adrian E},
  journal={Journal of the American Statistical Association},
  volume={90},
  number={430},
  pages={773--795},
  year={1995},
  publisher={Taylor \& Francis}
}

@article{lahiri2023sparse,
  title={Sparse-view cone beam {CT} reconstruction using data-consistent supervised and adversarial learning from scarce training data},
  author={Lahiri, Anish and Maliakal, Gabriel and Klasky, Marc L and Fessler, Jeffrey A and Ravishankar, Saiprasad},
  journal={IEEE Transactions on Computational Imaging},
  volume={9},
  pages={13--28},
  year={2023},
  publisher={IEEE}
}

@article{zheng2021concept,
  title={Concept, implementations and applications of Fourier ptychography},
  author={Zheng, Guoan and Shen, Cheng and Jiang, Shaowei and Song, Pengming and Yang, Changhuei},
  journal={Nature Reviews Physics},
  volume={3},
  number={3},
  pages={207--223},
  year={2021},
  publisher={Nature Publishing Group UK London}
}

@article{konda2020fourier,
  title={Fourier ptychography: current applications and future promises},
  author={Konda, Pavan Chandra and Loetgering, Lars and Zhou, Kevin C and Xu, Shiqi and Harvey, Andrew R and Horstmeyer, Roarke},
  journal={Optics Express},
  volume={28},
  number={7},
  pages={9603--9630},
  year={2020},
  publisher={Optical Society of America}
}

@article{li2025fourier,
  title={Fourier Ptychographic Microscopy with Iterative Weighted Side-window Guided Filtering},
  author={Li, Zewei and Lu, Dajiang and Tang, Qijian and Liu, Cheng and Tian, Yibin and Wu, Zongze},
  journal={IEEE Photonics Technology Letters},
  year={2026},
  volume={38},
  number={9},
  pages={600--603},
  publisher={IEEE}
}

@article{chetley2011estimation,
  title={Estimation of {CT} cone-beam geometry using a novel method insensitive to phantom fabrication inaccuracy: implications for isocenter localization accuracy},
  author={Chetley Ford, J and Zheng, Dandan and Williamson, Jeffrey F},
  journal={Medical Physics},
  volume={38},
  number={6Part1},
  pages={2829--2840},
  year={2011},
  publisher={Wiley Online Library}
}

@article{possolo2015simple,
  title={Simple guide for evaluating and expressing the uncertainty of NIST measurement results},
  author={Possolo, Antonio and others},
  journal={NIST Technical Note},
  volume={1900},
  year={2015}
}

@article{ly2017tutorial,
  title={A tutorial on Fisher information},
  author={Ly, Alexander and Marsman, Maarten and Verhagen, Josine and Grasman, Raoul PPP and Wagenmakers, Eric-Jan},
  journal={Journal of Mathematical Psychology},
  volume={80},
  pages={40--55},
  year={2017},
  publisher={Elsevier}
}

@article{okawa2026snr,
  title={{SNR} improvement in computed tomography-based distributed Brillouin sensing: a theoretical and numerical study},
  author={Okawa, Youhei and Furukawa, Hiromitsu},
  journal={Optics Express},
  volume={34},
  number={6},
  pages={11156--11170},
  year={2026},
  publisher={Optica Publishing Group}
}

@article{marcus2022information,
  title={Information content of and the ability to reconstruct dichroic X-ray tomography and laminography},
  author={Marcus, Matthew A},
  journal={Optics Express},
  volume={30},
  number={22},
  pages={39445--39465},
  year={2022},
  publisher={Optica Publishing Group}
}

@article{chai2026tsv,
  author  = {Chai, Shanglei and Zhang, Zhiyuan and Lu, Dajiang and Zhang, Chonglei and Zhang, Huan and Tian, Yibin},
  title   = {{Hy-SAN}: A Hybrid Shape-Aware Network for Artifact-Suppressed
Sparse-View Computed Tomography of Through-Silicon Vias},
  journal = {Nondestructive Testing and Evaluation},
  volume  = {},
  number  = {},
  pages   = {in press},
  year    = {2026}
}

@article{gao2010scatter,
  title={Scatter correction method for x-ray {CT} using primary modulation: phantom studies},
  author={Gao, Hewei and Fahrig, Rebecca and Bennett, N Robert and Sun, Mingshan and Star-Lack, Josh and Zhu, Lei},
  journal={Medical Physics},
  volume={37},
  number={2},
  pages={934--946},
  year={2010},
  publisher={Wiley Online Library}
}

@article{yang2026task,
  title={Task-driven lens design},
  author={Yang, Xinge and Fu, Qiang and Nie, Yunfeng and Heidrich, Wolfgang},
  journal={Optics Express},
  volume={34},
  number={5},
  pages={8961--8975},
  year={2026},
  publisher={Optica Publishing Group}
}

@inproceedings{souza2025latent,
  title={Latent space imaging},
  author={Souza, Matheus and Zheng, Yidan and Kang, Kaizhang and Mishra, Yogeshwar Nath and Fu, Qiang and Heidrich, Wolfgang},
  booktitle={IEEE/CVF Conference on Computer Vision and Pattern Recognition (CVPR)},
  pages={28295--28305},
  year={2025},
  organization={IEEE}
}






\clearpage

\appendix

\textbf{Supplemental Materials: Additional details of the parameters and perturbations used in the experiments}
\label{app:parameters}

\renewcommand{\thetable}{S\arabic{table}}
\setcounter{table}{0}

\begin{table*}[!h]
\centering
\caption{Nominal acquisition, reconstruction, and measurement estimation parameters used in the experiments. Values denote the reference configuration unless otherwise stated.}
\label{tab:parameter}

\small
\setlength{\tabcolsep}{3pt}

\begin{tabular}{
>{\centering\arraybackslash}p{1cm}
p{4cm}
p{4cm}
p{6.5cm}}
\toprule
Cat & Parameter & Nominal value / range & Description \\
\midrule

\multirow{6}{*}{
\rotatebox[origin=c]{90}{Object/task}
}
& Lateral ROI
& $4.8\times4.8~\mathrm{mm}^2$
& Canonical layered phantom \\

& Object thickness
& $3.2~\mathrm{mm}$
& Canonical layered phantom \\

& Axial gap (ATG), $L$
& $120~\mu\mathrm{m}$
& Reference plane to feature termination\\

& Reference plane, $Z_{\mathrm{ref}}$
& $0.55~\mathrm{mm}$
& Axial location of the Cu reference layer \\

& Cu layer thickness
& $35~\mu\mathrm{m}$
& Thickness of the Cu reference layer \\

& Material classes
& Cu and FR4
& Two effective PCB material classes \\

\midrule

\multirow{12}{*}{
\rotatebox[origin=c]{90}{Acquisition}
}
& Number of views, $N_v$
& $12,18,24,32,48,64$
& Sparse-view settings \\

& SOD / SDD
& $120 / 360~\mathrm{mm}$
& Source-object / Source-detector Distance  \\

& Magnification, $M_0$
& $3.0$
& $M_0=\mathrm{SDD}/\mathrm{SOD}$. \\

& Obliquity angle, $\alpha$
& $35^\circ$
& Between source direction and the PCB normal \\

& Starting azimuth
& $0^\circ$
& Initial source position on the trajectory \\

& Detector pixel pitch
& $0.100~\mathrm{mm}$
& Physical detector sampling pitch. \\

& Detector dimensions
& $256\times256$
& Enlarged to guarantee coverage if needed \\

& Photon budget, $N_0$
& $2.5\times10^5$ /pixel
& Nominal incident photon budget \\

& Photon budget sweep
& $2\times10^4$--$10^6$ /pixel
& Range to evaluate noise and the radiometric-stage marginalized CRLB. \\

& Detector PSF, $\sigma_b$
& $0.65$ pixel
& Gaussian detector-blur SD \\

& Read-noise SD, $\sigma_r$
& $8~e^{-}$
& SD of additive Gaussian read noise \\

& Flat-field average, $N_F$
& $8$
& Number of open-beam frames averaged  \\

\midrule

\multirow{7}{*}{
\rotatebox[origin=c]{90}{
\parbox{2.5cm}{\centering Spectrum/ detector}
}
}
& Tube voltage, $V_{\mathrm{kVp}}$
& $120~\mathrm{kVp}$
& Endpoint voltage of the X-ray spectrum \\

& Min photon energy
& $15~\mathrm{keV}$
& Lowest photon energy in the spectrum \\

& Number of energy bins, $N_E$
& $48$
& Number of discrete spectral samples in the polychromatic forward model \\

& Spectrum model
& Kramers-type approximation
& Numerical approximation for the source spectrum \\

& Detector response
& $\propto E$
& Effective energy-dependent weighting \\

& Al filtration
& $0.5~\mathrm{mm}$
& Aluminum filtration applied to the source \\

& Cu filtration, $t_{\mathrm{Cu}}$
& $0.20~\mathrm{mm}$
& Cu filtration and the reference value for the Stage-1 spectral perturbation. \\

\midrule

\multirow{7}{*}{
\rotatebox[origin=c]{90}{Reconstruction}
}
& Computation
& ASTRA CUDA
& GPU-based projection and reconstruction \\

& Baseline
& SIRT + 3D TV
& Main baseline for comparison \\

& SIRT outer iterations
& $5$
& Outer regularization cycles \\

& SIRT iterations per outer cycle
& $4$
& Algebraic reconstruction iterations within each outer cycle. \\

& TV regularization weight
& $0.010$
& Weight of the 3D TV regularizer \\

& TV iterations
& $12$
& TV iterations in each reconstruction \\

& TIGRE baseline
& SART, 12 iterations
& Secondary reconstruction used only when geometry mapping is enabled. \\

\bottomrule
\end{tabular}
\end{table*}

\begin{table*}[!t]
\centering
\caption{Physical-perturbation parameters used in the experiments. Values denote the reference configuration unless otherwise stated.}
\label{tab:perturbation}
\small

\begin{tabular}{
>{\centering\arraybackslash}p{1cm}
p{3.5cm}
p{4cm}
p{6.5cm}}
\toprule
Cat & Parameter & Nominal value / range & Description \\
\midrule

\multirow{4}{*}{%
\rotatebox[origin=c]{90}{%
\parbox{3.0cm}{\centering Radiometric}%
}}
& Scatter fraction, $s$
& $0$ nominal; $0.05$ perturbation
& Amplitude of the low-frequency additive scatter component \\

& Effective blur, $\sigma_b$
& $0.65$ px nominal; $1.10$ px perturbation
& Effective combined detector/focal-spot blur \\

& Tube voltage, $V_{\mathrm{kVp}}$
& $120$ kVp nominal; $110$ kVp perturbation
& Effective spectral-endpoint perturbation \\

& Cu filtration, $t_{\mathrm{Cu}}$
& $0.20$ mm nominal; $0.30$ mm perturbation
& Effective filtration perturbation used to represent spectral perturbation. \\

\midrule

\multirow{4}{*}{%
\rotatebox[origin=c]{90}{%
\parbox{2.7cm}{\centering Geometric}%
}}

& Relative magnification, $\delta_M$
& $0$ nominal
& Global magnification perturbation 
$M=M_0(1+\delta_M)$. \\

& Global perturbation, $\delta_M$
& $\approx-5.81\times10^{-3}$
& Effective relative magnification change \\

& Obliquity offset, $\Delta\alpha$
& $0$ nominal; $+0.30^\circ$ perturbation
& Common angular calibration perturbation applied to the acquisition trajectory \\

& Detector-$v$ shift, $\Delta v$
& $0$ nominal; $0.12~\mathrm{mm}$ perturbation
& Global detector translation along the local detector-$v$ direction. \\

\midrule

\multirow{4}{*}{%
\rotatebox[origin=c]{90}{%
\parbox{1.9cm}{\centering Warping}%
}}

& Overall warp amplitude, $a_w$
& $0$ nominal; $120~\mu\mathrm{m}$ perturbation
& Overall magnitude of the PCB deformation \\

& Bow coefficient, $a_b$
& $0.65$
& Relative weight of the quadratic bow\\

& Saddle coefficient, $a_t$
& $0.35$
& Relative weight of the deformation $X_nY_n$. \\

& Piston constant, $c$
& $0.60$
& Constant offset to remove the piston component
of the quadratic bow mode. \\

\midrule

\multirow{3}{*}{%
\rotatebox[origin=c]{90}{%
\parbox{1.8cm}{\centering Selection}%
}}

& BIC improvement threshold
& $\Delta\mathrm{BIC}\ge10$
& Minimum improvement required before accepting a more complex perturbation model \\

& Geometry jitter
& $0$--$75~\mu\mathrm{m}$
& View-dependent geometric perturbation \\

& Unmodeled-jitter handling
& Not explicitly estimated
& Random per-view jitter remains outside the fitted low-dimensional perturbation family \\

\bottomrule
\end{tabular}
\end{table*}

\end{document}